\documentclass[12pt,preprint]{aastex}
\usepackage[colorlinks,citecolor=blue]{hyperref}

\title{VLBI studies of DAGN and SMBHB hosting galaxies}

\author{T. An$^*$\thanks{antao@shao.ac.cn}, P. Mohan$^*$\thanks{pmohan@shao.ac.cn}, S. Frey$^\dagger$\thanks{frey.sandor@csfk.mta.hu}\\
$^*$: Shanghai Astronomical Observatory, 80 Nandan Road, Xuhui District, Shanghai 200030, China.\\
$^\dagger$: Konkoly Observatory, MTA Research Centre for Astronomy and Earth Sciences, Konkoly Thege Mikl\'os \'ut 15-17, H-1121 Budapest, Hungary.}

\begin{abstract}
Dual active galactic nuclei (DAGN) and supermassive black hole binaries (SMBHBs) at kpc and pc-scale separations, respectively, are expected during stages of galaxy merger and evolution. Their observational identification can address a range of areas of current astrophysics frontiers including the final parsec problem and their contribution towards the emission of low-frequency gravitational waves. This has however been difficult to achieve with current spectroscopy and time domain strategies. Very long baseline interferometry (VLBI) as a method of directly imaging radio structures with milli-arcsecond (mas) and sub-mas resolutions is introduced as a possible means of detecting DAGN and SMBHBs. We motivate its usage with expected observational signatures and cite some studies from literature to illustrate its current status, and present an updated list of candidates imaged with high-resolution radio observations. We then recall some shortcomings of the method with possible solutions and discuss future directions, relevant to large surveys with the upcoming Square Kilometer Array and future space VLBI missions.\\
{\bf Key points}
\begin{itemize}
\item Dual active galactic nuclei (DAGN) and supermassive black hole binaries (SMBHBs): observational status.
\item Very long baseline interferometry (VLBI): a promising tool to detect dual AGN and SMBHBs. 
\item Observation strategies for the future.
\end{itemize}  
\end{abstract}

\begin{document}

\maketitle

\section{Introduction and motivations}

Hierarchical models of cosmological structure formation \citep[e.g.][]{2005Natur.435..629S} expect major mergers shaping galaxy evolution. Binary evolution upto kpc scales are governed by dynamical friction \citep{1943ApJ....97..255C} which here involves a decreasing binary separation through the extraction of their effective angular momentum and energy through stellar encounters \cite[e.g.][]{2001ApJ...563...34M,2005LRR.....8....8M} over $<$ 10 Myr timescales, which can be sped up by gravitational encounters with intervening gas \cite[e.g.][]{2007Sci...316.1874M,2007MNRAS.379..956D,2017MNRAS.469.4258T}; the kpc-separated central objects appear as a dual AGN (DAGN) \cite[e.g.][]{2003ApJ...596..860M,2005LRR.....8....8M}, possibly owing to rapid accretion. A continued merger aided by stellar gravitational interactions leads to a pc-scale separation  Keplerian bound supermassive black hole binary (SMBHB) \citep[e.g.][]{1980Natur.287..307B}. The SMBHB is expected to be strongly gravitationally bound when its specific binding energy exceeds the nuclear stellar velocity dispersion based energy with a fiducial separation of $\sim 2.7$ pc marking this transition assuming a SMBH mass (smaller mass binary companion) of $10^8 M_\odot$ and a stellar bulge velocity dispersion of 200 km s$^{-1}$ \citep{2005LRR.....8....8M}. The central core is now depleted of stars and gas rendering dynamic friction ineffective in enabling further merger, the `final parsec problem' \cite[e.g.][]{1980Natur.287..307B}. This could be overcome through three-body interactions involving a companion massive black hole \cite[e.g.][]{2016MNRAS.461.4419B}, if the initial extraction of angular momentum and energy proceeds through episodes of stellar and gas interactions \cite[e.g.][]{2017MNRAS.472..514G}, or if the stellar loss cone can be re-populated efficiently thus enabling a continuity in stellar encounters \cite[e.g.][]{2017MNRAS.464.2301G}. These cause the bound SMBHB to enter the gravitational wave regime and tend towards coalescence rapidly. Alternatively, simulations of mergers in triaxial rotating galaxies indicate that the binary hardening rates are sufficient to enable efficient coalescence, avoiding stalling of the merger at pc-scales \citep{2006ApJ...642L..21B}.

Observational identification of DAGN and SMBHBs, however, has mostly been indirect and serendipitous so far,  mainly owing to the small separation and associated physical processes involved in both being distinct, thus requiring different search strategies in these classes of objects. Employed strategies have mainly involved spectroscopy and timing based on the double-peaked line profiles and continuum light curves from candidate sources. A range of multi-wavelength searches 
have been met with varying levels of success \citep[e.g.][]{2003ApJ...582L..15K,2006MmSAI..77..733K,2012ApJ...746L..22K,2013ApJ...777...64C}. Double peaked spectral emission lines (e.g. H$\beta$, [O III]) separated in radial velocity by $\sim$ few hundred -- thousand km s$^{-1}$ are expected in a merging system hosting narrow- and broad-line regions   (at larger and smaller physical distances from the central engine, respectively) \cite[e.g.][]{2007ApJ...660L..23G,2009ApJ...705L..20X}. Double peaked narrow lines are expected to indicate a DAGN while broader peaked lines could indicate a SMBHB. Though, these lines could originate from the superposition of unrelated overlapping AGN, peculiar geometry of the narrow line region (NLR) and kinematics within a single cloud (e.g. clumpy structure), powerful biconical outflows, jet--NLR interaction \citep[e.g.][]{2009ApJ...705L..20X,2012ApJ...753...42C} or as equal strength peaks from a rotating disk or ring ionized by a single underlying source \cite[e.g.][]{2012ApJ...752...63S}. 
 In fact, only a tiny fraction of AGN showing double-peaked emission line profiles have been identified as DAGN \citep[e.g.][]{2011ApJ...735...48S,2012ApJ...745...67F,2015ApJ...813..103M}. Updated lists of inferred DAGN compiled from the literature are presented in \citet{2018JApA...39....8R} and \citet{2018BSRSL..87..299D}. The continuum or line-based optical light curves may indicate quasi-periodicity over timescales of a few to tens of years, possibly related to the orbital period   of a Keplerian SMBHB \citep[e.g.][]{
2015MNRAS.453.1562G,
2016MNRAS.463.2145C,2016MNRAS.463.1812M}. The well studied AGN OJ 287 putatively hosts a SMBHB, inferred through a pair of periodic optical outbursts every $\sim$ 12 yr \cite[e.g.][]{1988ApJ...325..628S,2011ApJ...729...33V} resulting from the impinging of the accretion disk of the primary by the secondary SMBH at these intervals during its orbital motion \cite[e.g.][]{1996ApJ...460..207L}. However, the recent study of \cite{2018MNRAS.478.3199B} notes that jet precession can explain the periodic variability though the driver of precession may still be a SMBBH.

As spectroscopic and timing searches are indirect, they require ruling out competing models and need independent confirmation. 
High resolution X-ray imaging spectroscopy observations with Chandra have resulted in the discovery of DAGNs \citep[e.g. NGC 6240:][]{2003ApJ...582L..15K}, \citep[Mrk 463:][]{2008MNRAS.386..105B}, \citep[Mrk 739:][]{2011ApJ...735L..42K}, \citep[Mrk 266:][]{2012AJ....144..125M}, and others from studies of candidate samples \citep[e.g.][]{2012ApJ...746L..22K,2013ApJ...777...64C}.
Similar targeted X-ray and optical imaging observations are often hampered by obscuration and scattering of the central kpc-scale region by interstellar gas in the host galaxy or unsuitable orientation of the galaxy. Further, there is not sufficient resolution to clearly separate the two nuclei in these images, especially in objects at cosmological distances (well beyond the local Universe). Very long baseline interferometry (VLBI) was developed during the 1960s to resolve fine structures of compact bright radio sources, and is the highest-resolution imaging technique \citep{2001ARA&A..39..457K}; VLBI observations with baselines $\sim$ few thousand kilometers can resolve milli-arcsecond (mas) structures (physical sizes of pc to kpc) and directly image even closely-separated nuclei, thus offering a promising method of spatially identifying DAGN and SMBHBs. In comparison, the angular resolutions achievable in other wavebands are $\sim$ 0.5 arcsec with the {\em Chandra} X-ray Observatory, and $\sim$ 0.1 arcsec with the {\em Hubble Space Telescope} in the optical and ultraviolet and with the {\em W. M. Keck Observatory} adaptive optics system in the near infrared. The VLBI radio observations are then indispensable in studying binary evolution from kpc to pc scales \citep[e.g.][]{2017MNRAS.471.3646P} and constructing a statistically viable sample through direct imaging observations, clarifying mechanisms enabling coalescence of a binary before gravitational wave emission becomes dominant (to avoid or overcome the final parsec problem), and helping to understand the nano-Hz gravitational wave background emission \citep[e.g.][]{2017NatAs...1..886M} using pulsar timing arrays \cite[e.g.][]{2008MNRAS.390..192S,2010CQGra..27h4013H}. 

The method and observational signatures relevant to it are discussed in Section \ref{obs}. 
Future directions relevant to VLBI observation of DAGN and SMBHBs are then discussed in Section \ref{future}.



\section{VLBI observations of DAGN and SMBHBs}
\label{obs}

For a source at $z$ = 0.1, 0.5 and 1, assuming a standard $\Lambda$CDM cosmology ($H_0$ = 71 km s$^{-1}$ Mpc $^{-1}$, $\Omega_M$ = 0.27 and $\Omega_V$ = 0.73) and a binary separation $R$ = 1 pc - 1 kpc, the required angular resolution $R/D_A$ (where $D_A$ is the angular diameter distance) is between 0.12 mas and 0.54 arcsecond, and is summarized in Table \ref{zres}. The typical angular resolutions of major VLBI networks are 0.26 - 0.5 mas at the 22 GHz frequency band \citep[e.g.][]{2018NatAs...2..118A}, suitable for imaging most nearby DAGN and SMBHBs. 

\begin{table}
\centerline{
\begin{tabular}{cccc}\hline
$z$ & $D_A$ & $R$ & $R/D_A$ \\
    & (Gpc) & (pc) & (mas)  \\ \hline \hline
0.1 & 0.38  & 1    & 0.54 \\
    &       & 1000 & 542.8 \\
0.5 & 1.25  & 1    & 0.17 \\
    &       & 1000 & 165.0 \\
1.0 & 1.66  & 1    & 0.12   \\
    &       & 1000 & 124.3 \\ \hline
    \end{tabular}}
\caption{Required resolution to clearly identify DAGN and SMBHB in the relatively nearby universe ($z \leq 1$).}
\label{zres}
\end{table}

Prominent signatures of DAGN and SMBHBs directly relevant to VLBI include double compact flat- or inverted-spectrum components, i.e., double cores, merger remnants   of two active star forming nuclei and their influence on circumnuclear star formation \citep{2018BSRSL..87..299D} and the presence of peculiar jet morphologies including S, Z and X-shapes, and helical pc-kpc scale structures \citep[e.g.][]{2006MmSAI..77..733K} which can arise due to tidal interaction of a binary SMBH with the accretion disk or jet base causing a precession \cite[e.g.][]{1980Natur.287..307B} or through a flip in the black hole spin during a merger which can re-orient the associated emergent jet \citep{2002Sci...297.1310M}.

With a small field of view, VLBI has not been employed as a powerful searching technique, but its unique high resolution yields stringent evidence to confirm or reject a candidate DAGN or SMBHB identified by other methods. Observations have thus far resulted in the detection of a SMBHB with a separation of $\sim$ 7 pc in 0402$+$379 \citep{2006ApJ...646...49R}, confirmed with follow-up observations after $\sim$ 12 yr \citep{2017ApJ...843...14B} which additionally help constrain the jet radiative properties and the orbital velocity, period and black hole mass of the system. A promising candidate with a sub-pc separation was reported in NGC 7674 \citep{2017NatAs...1..727K}, and requires follow-up observations (multi-epoch and multi-frequency VLBI) to confirm the detection of double radio cores. The imaging of NGC 5252 has been employed in identifying a possible off-nucleus ultra-luminous X-ray (ULX) source as a low luminosity AGN, thus making it a candidate DAGN \citep{2017MNRAS.464L..70Y}. The study of follow-up VLBI observations of NGC 5252 \citep{2018MNRAS.480L..74M} indicates either a ULX hosting an intermediate-mass black hole or a low-mass AGN. Based on optical periodic variability over a timescale of $\sim$ 5 yr, the source PG 1302$-$102 was proposed as a SMBHB candidate \citep{2015Natur.518...74G}. In addition to multi-wavelength periodicity \citep{2015Natur.525..351D}, PG 1302$-$102 exhibits interesting pc-scale jet structure, with \cite{2015MNRAS.454.1290K} and \cite{2016MNRAS.463.1812M} exploring its helical nature, and \cite{2018A&A...615A.123Q} applying a model of a precessing jet nozzle to infer the association between the pc-scale radio jet and the optical variability timescale.

Early VLBI searches based on candidate double-peaked narrow optical emission lines (DPNL) AGN did not detect double cores \citep[e.g.][]{2011AJ....141..174T}, possibly due to limited sensitivity or as DPNL AGN may not host SMBHBs. Some recent targeted VLBI observations of DPNL AGN, with a typical sensitivity of 20 $\mu$Jy, show mixed results: inference of double cores in SDSS J1502$+$1115 \citep{2014Natur.511...57D} and SDSS J1536$+$0441 \citep{2010ApJ...714L.271B}, only a single radio-emitting AGN detected in SDSS J1425$+$3231 \citep{2017IAUS..324..223G}, 3C 316 \citep{2013MNRAS.433.1161A}, and some others \citep{2014MNRAS.443.1509G,2016ApJ...826..106G}, posing an interesting question of whether both SMBHs become active in radio bands. The study of \cite{2011MNRAS.410.2113B} involves a blind search for sources hosting dual radio bright components in archival multi-frequency VLBI images, finding a general paucity of resolved pc-scale stalled binaries suggesting that they likely exist within a common bright radio envelope and coalescence is rapid. An updated but non-exhaustive compilation of VLBI observed candidate DAGN and SMBHBs is presented in Table \ref{comp}. 


\begin{table}
\begin{tabular}{lllll}\hline
Source \&                 & Radio & Redshift & $\nu$: $S_{\nu,{\rm VLBI}}$           & Core Sep. \\
attributes                & Morphology   & $z$        & (GHz: mJy)               & (kpc) \\ \hline \hline
NGC 5252$^{(1),F}$ & Double compact cores & 0.022 & 1.6: 1.80, 3.60 & 10.0 \\
3C 75$^{(2)F}$        & Double components        & 0.023     & 1.4: 111.0, 41.0        & 7.0 \\
                        &                   &           & 2.4: 63.0, 22.0        & \\
                         &                   &           & 4.8: 38.0, 13.0         & \\
                        &                   &           & 8.6: 4.0, 5.0         & \\
NGC 6240$^{(3,4)F}$      & One compact core + SNR & 0.024 & 1.6: 2.09, 3.95 & 1.4 \\
                         &                                 &       & 5.0: 2.6, 6.0   &     \\
NGC 7674$^{(5)I}$      & Double compact cores & 0.030 & 15: 0.9, 0.9          & 3.5$\times 10^{-4}$ \\
NGC 326$^{(6)I}$    & Double components         & 0.047     & 1.4: 5.33, 1.16       & 4.8 \\
                       &                    &           & 4.8: 10.42, 0.78      & \\ 
0402$+$379$^{(7)I}$    & MSO + naked core         & 0.055    & 5.0: 53.2, 10.7 & 7.3$\times 10^{-3}$\\
                         &                    &          & 8.0: 60.5, 15.3 &                    \\
                         &                    &          & 15.0: 52.4, 14.8 &                    \\
                         &                    &          & 22.0: 35.3, 10.7 &                    \\
                         &                    &          & 43.0: 23.4, 8.0 &                    \\
PKS 1155$+$251$^{(8)F}$ & CSO       & 0.203    & 24: 2.74, 147.60         & 1.2$\times 10^{-2}$ \\
                          &                   &          & 43: 3.37, 118.87         &  \\
J1536$+$0441$^{(9,10)S}$ & Double compact cores        & 0.390    & 5.0: 0.72, 0.27          & 5.1 \\
                         &                    &          & 8.0: 1.17, 0.24          & \\
                         &                    &          & 8.5: 1.17, 0.27         &  \\
J1502$+$1115$^{(11)S}$& Double components        & 0.390     & 1.4: 2.58, 7.04         & 7.4 \\
                         &                   &           & 5.0: 0.93, 2.33         & \\
                         &                   &           & 8.5: 0.61, 1.28         & \\
J1502$+$1115 S$^{(12)F}$ & Double components    & 0.390     & 1.7: 0.95, 0.92         & 1.4$\times 10^{-1}$ \\
                            &                &           & 5.0: 0.86, 0.87         & \\
J1425$+$3231$^{(13)F/S}$& Double components & 0.478    & 1.7: 0.46, 0.23          & 2.6  \\
                          &                   &          & 5.0: 0.35, $<$ 0.09      & \\
3C 316$^{(14)S}$         & Complex jet       & 0.580    & 1.6: 26.2, 16.5          & 1.8$\times 10^{-1}$ \\ \hline
\end{tabular}
\caption{List of candidate and confirmed VLBI observed DAGN and SMBHB hosts. Attributes relating to spectral type include $F$: flat, $S$: steep, $F/S$: flat and steep, and $I$: inverted. References: (1): \citet{2017MNRAS.464L..70Y}, (2): \citet{2004MNRAS.351..101K}, (3): \citet{2003ApJ...582L..15K}, (4): \citet{2011AJ....142...17H}, (5): \citet{2017NatAs...1..727K}, (6): \citet{2001A&A...380..102M}, (7): \citet{2006ApJ...646...49R}, (8): \citet{2017MNRAS.471.1873Y}, (9): \citet{2010ApJ...714L.271B}, (10): \citet{2009ApJ...699L..22W}, (11): \citet{2011ApJ...740L..44F}, (12): \citet{2014Natur.511...57D}, (13): \citet{2012MNRAS.425.1185F}, (14): \citet{2013MNRAS.433.1161A}. } 
\label{comp}
\end{table}


Some general inferences can be drawn from the existing VLBI detection of DAGN and SMBHBs. 
\begin{itemize}
\item A stringent requirement is that both nuclei are active radio emitters. 
  This may be the case especially in the SMBHB scenario as they may exist within a common radio bright core \cite[e.g.][]{2011MNRAS.410.2113B}, thus necessitating high resolution follow-up and monitoring.
\item   If a fraction of DAGN are radio loud/quiet or quiet/quiet pairs \citep[e.g., J1643$+$3156][]{2011ApJ...736..125K} either owing to unbeamed emission from a constituent, or the quenching of jet production due to destruction of the primary BH accretion disk by the secondary BH, a blind VLBI search for radio pairs may then fail to identify them, biased towards single radio-loud AGN hosts.
\item The radio flux densities are typically at mJy and sub-mJy levels, requiring high sensitivity VLBI imaging. 
\item Most candidates   found in radio so far are DAGN. To discover sub-pc scale SMBHBs (likely sources of low-frequency gravitational waves) by direct imaging, the expected sub-mas resolution with current mm-wavelength VLBI or future space VLBI at centimeter wavelengths would be necessary. The existing VLBA can achieve resolutions of 0.07 mas and 0.18 mas at 86 and 43 GHz, respectively, enabling the resolving of DAGN upto moderate redshifts. For those with steep-spectrum cores, VLBI at lower radio frequency (e.g., $\leq$ 5 GHz) would be suitable; that has to involve space VLBI to reach sub-mas resolution \cite[0.4 mas resolution at 1.67 GHz with a baseline of 100,000 km, e.g.][]{2018arXiv180810636A}.
\item The radio activity appears to be recently triggered (by galaxy merger) and can be characterized as systems with high accretion rates. The resulting DAGN and SMBHBs resemble compact symmetric objects (CSOs) or compact steep-spectrum (CSS) sources   , both of which are classified as young radio galaxies. Typical examples include the inverted-spectrum cores in 0402$+$379 \citep{2006ApJ...646...49R}, and in NGC 6240 \citep{2011AJ....142...17H}. Further, as their core flux density only accounts for a small fraction of the total flux density, this indicates that their radio emission is dominated by extended jets and lobes. 
\end{itemize}

\section{Discussion and future directions}
\label{future}

The expected number of binary SMBH hosts can be inferred from the AGN luminosity function by accounting for the rate of mergers, the timescale of existence of the AGN phase, the SMBH mass function, and the spatial distribution of AGN in the redshift volume being probed \citep[e.g.][]{2009ApJ...700.1952H}; though, estimates are model dependent and hence could result in large uncertainties. Modeling SMBHBs during the inspiral phase driven by gas and gravitational wave emission, \cite{2018ApJ...863..185D} estimate $\sim$ 100 SMBHBs at $z \leq 0.5$ which can show periodic variability $<$ 10 yr that can be resolved using mm-wavelength VLBI, relevant to facilities such as the Event Horizon Telescope. The study of \cite{2015MNRAS.453.1562G} predicts 450 candidates showing optical periodic variability (assuming $R \leq 0.01$ pc, limiting $V$ magnitude of 20 and $z \leq 4.5$) but find only 111 candidates from a sample of 243500 quasars, indicating a conservative agreement. A similar $\geq$ 100 candidate DAGN ($R \sim$ kpc) in high frequency radio (e.g. 1.6 and 5 GHz) surveys with either medium sensitivity $\sim$ 0.1 mJy and sky area of $\sim$ thousand square degree or high sensitivity $\sim$ 10 $\mu$Jy and sky area of $\sim$ few hundred square degree   is predicted by \cite{2014arXiv1402.0548B}. The study of \cite{2016MNRAS.463.2145C} finds a similarly conservative estimate of 33 candidates indicating periodic optical variability from a sample of 35383 quasars. SMBHBs indicate a transition to the gravitational wave emitting regime and the study of \cite{2018ApJ...856...42S} characterizes the expected gravitational wave background limits from pulsar timing arrays and finds a moderate tension with the above observational estimates, suggesting that the inferred candidate samples are contaminated with false detections. The above estimates based on periodic variability indicate an efficiency of 0.05 -- 0.09 \% which may only be upper limits, but are likely to improve with continued monitoring which will enable a more robust statistical identification. With the Square Kilometre Array (SKA) phase 1 Mid configuration in the 4 cm band ($\sim$ 8.4 GHz; applicable to possible SKA-VLBI observation) and a maximum baseline of $\sim$ 10000 km (between SKA 1-Mid and European VLBI stations), the resolution is about 0.15 mas. The expected image sensitivity is 3 $\mu$Jy/beam for an integration time of 1 hour with the full SKA1-Mid in the global VLBI network \citep{2015aska.confE.143P} so that the combination of high resolution and high sensitivity with reasonable integration time can resolve sub-mas structures in DAGN and SMBHB candidates \citep{2015aska.confE.151D}. 

The major shortcoming of the current VLBI technique is the small field of view at a given time, limiting its applicability for blind surveys. Further issue in the observation itself may arise from a combination of selection criteria and biases (methodology), instrumental properties and limitations (detection sensitivity) and intrinsic source-based properties (accretion and jet production mechanisms, radio quietness and non-active galaxies) \citep[e.g.][]{
2018arXiv180302831S}.   Further, as only $\sim$ 10\% AGN are typically radio loud, fainter sources should be sufficiently radio bright (and compact) to be VLBI detected; more DAGN are expected to be detected with an increasing VLBI sensitivity. Continuing efforts towards addressing these include the upgrading of equipment with higher data recording rate and updated software correlators, developing robust and time optimal algorithms for data handling and reduction, and their deployment and testing on smaller arrays and scaling up \citep[e.g.][]{2018NatAs...2..118A}. 
For faint and un-beamed AGN hosting SMBHBs (non-existent or weak large-scale jet), the multi-beam, wide-field SKA can offer fast surveys, and its high sensitivity allows for discovering even weak radio loud/quiet pairs. It is essential to thus address why a larger number of DAGN and SMBHBs are yet to be observed using current strategies.
An alternate means to resolve pc and sub-pc separation SMBHBs is by resorting to space VLBI observations with two or three space-based telescopes, operational in the 1.5 - 8.4 GHz frequency range and offering few tens of $\mu$as-mas resolutions. The space VLBI is expected to resolve the central regions of nearby bright AGN \citep[e.g. 3C 84, ][]{2018NatAs...2..472G} and revealing the precession of the jet nozzle,  and to search for sub-pc-separation SMBHBs in elliptical galaxies which have likely experienced multiple mergers. Such space VLBI telescopes can be connected with large ground-based telescopes, such as SKA1-Mid, FAST 500m, Arecibo 300m, Effelsberg 100m and Green Bank 110m, pushing detection limits down to sub-mJy levels.   Additionally, one can use ground based mm-wavelength facilities to achieve sub-mas resolutions \cite[e.g.][]{2011ApJ...735...57B} and even approach mJy sensitivity in the near future.

\section*{Acknowledgements}
We thank the anonymous reviewers for the comments and suggestions which helped improved our paper. This work is supported by the International Key R\&D Program of China (grant number 2018YFA0404603). PM is supported by the National Science Foundation of China (grant no. 11650110438). SF thanks the Hungarian National Research, Development, and Innovation Office (OTKA NN110333) for support.  We acknowledge the use of the Cosmology Calculator \citep{2006PASP..118.1711W} to estimate data presented in Table 1. The data presented in Table 2 is compiled from literature with the appropriate source cited, and is intended in identifying diverse and similar characteristics of DAGN and SMBHB systems relevant to VLBI radio observations.

\end{document}